# Interval State Estimation with Uncertainty of Distributed Generation and Line Parameters in Unbalanced Distribution Systems

Ying Zhang, *Student Member, IEEE*, Jianhui Wang, *Senior Member, IEEE*, Zhengshuo Li, *Member, IEEE*

***Abstract*— Distribution system state estimation (DSSE), which provides critical information for system monitoring and control, is being challenged by multiple sources of uncertainties such as random meter errors, stochastic power output of distributed generation (DG), and imprecise network parameters. This paper originally proposes a general interval state estimation (ISE) model to simultaneously formulate these uncertainties in unbalanced distribution systems by interval arithmetic. Moreover, this model can accommodate partially available measurements of DG outputs and inaccurate line parameters. Further, a modified Krawczyk-operator (MKO) algorithm is proposed to solve the general ISE model efficiently, and effectively provides the upper and lower bounds of state variables under coordinated impacts of these uncertainties. The proposed algorithm is tested on unbalanced IEEE 13-bus and 123-bus systems. Comparison with various methods including Monte Carlo simulation indicates that the proposed algorithm is many orders of magnitude faster and encloses tighter boundaries of state variables.**

## I. Introduction

DISTRIBUTION system state estimation (DSSE) converts redundant meter readings and other available information into an estimate of system states and thus develops into a subject of active research [1], [2]. This subject is largely driven by the diffusion of distributed generation (DG) such as wind turbine generators (WTGs) and photovoltaic (PV) panels. DG has advantages of low investment costs, flexible and eco-friendly operations, and low power losses [3]. However, the variability and intermittency of DG pose significant uncertainty to DSSE [4]. Apart from the uncertainty from these emerging DG units, inputs to DSSE also contain measurements with noises and imprecise line parameters. For example, the uncertainty of line parameters originates from varying field ambient conditions and aging wirings. While the uncertainty of measurements is commonplace and their impacts on the DSSE are investigated in [5], multiple other uncertainties besides DG uncertainty in DSSE call for innovative solutions.

Uncertainty studies that account for the variable and stochastic nature of input data in conventional DSSE are pursued by classical Monte Carlo (MC) simulation, as reviewed in [1]. In these studies, DG power outputs are assumed available in real time via measuring instruments installed by distribution system operators (DSOs) [6] or in the form of pseudo-measurements that follow Gaussian distributions [7]. Nevertheless, these assumptions might be impractical due to 1) currently limited metering and communication infrastructure, and 2) lack of specific agreements between DG operators and DSOs [8]. Moreover, the stochastic nature of DG outputs weakens the assumption that the DG outputs follow a known family of parametric distributions [9]. As discussed in [10]–[17], the statistical data of DG outputs are a prerequisite in MC simulation, however, such information may not be available in practice. In addition, these methods based on MC simulation require plenty of runs for various combinations of measurement samplings and/or DG outputs and thus are generally applied for evaluating the overall accuracy of state estimators [5].

Motivated by the deficiencies of these methods, interval state estimation (ISE) is proposed to obtain the boundaries of state variables, which provide more intuitive information such as the upper and lower bounds of these states [10]. In ISE, all data with uncertainty are modeled as inputs in the interval form, since the upper and lower limits are available in most practical cases. For example, the range of line parameters can be specified (e.g., within ± 5% of their nominal values) [17].

Boundary optimization methods such as [10]–[13] are proposed to address the ISE problem by maximizing and minimizing the state variables that meet all constraints from measurements. For instance, in [10], a constrained nonlinear programming approach using the measurements from supervisory control and data acquisition (SCADA) systems is used to obtain the ranges of states in transmission systems. The solving method is applied to distribution systems in [13]. However, with more phasor measurement units (PMUs) or micro-PMUs emerging at the distribution level, this approach cannot deal with hybrid measurements including PMU data. Moreover, the authors of [13] did not consider the uncertainty of line parameters. In addition, the limitation of these optimization-based methods is that lower and upper bounds of each state variable need to be computed separately, and thus the total number of the optimization models for all states proliferates with the scale of the distribution systems. This leads to their low efficiency in the solving process.

Recent efforts to apply interval arithmetic to study uncertainties in power system operations are noteworthy, such as power flow calculation in [14] and [15], reliability evaluation in [16], and ISE in [17] and [18]. In ISE, interval arithmetic deals with the uncertain inputs that lie within a certain interval and enables the direct computation towards the bounds of state variables. For instance, focusing on transmission systems, an ISE model with only PMU measurements is formulated as interval linear equations in [17], and the ranges of states with

Y. Zhang, J. Wang, and Z. Li are with the Department of Electrical and Computer Engineering, Southern Methodist University, Dallas, TX, USA (e-mail: yzhang1@smu.edu; jianhui@smu.edu; zhengshuol@smu.edu).

the line parameter uncertainty are solved. However, such a high PMU deployment is not available at the distribution level, and the impacts of DG uncertainty are not considered. Further research is conducted in active distribution systems. In [18], an iterative Krawczyk-operator algorithm is used to obtain interval states, which takes the solution solved by interval Gaussian elimination (IGE) as initial values of the states. Nevertheless, IGE presents the drawback of "wrapping effect", where the widths of intervals expand since each interval variable is treated independently. When IGE is applied to distribution systems with high uncertainty, this over-conservatism is further intensified. As a result, the IGE-based Krawczyk operator (IKO) is computationally expensive to obtain final states since the initial states are far from them [13]. In addition, all DG outputs in [18] are assumed to obey Gaussian distributions. As mentioned above, this assumption may not be practical.

To sum up, the existing studies still lack generality in modeling to formulate multiple uncertainties, and most of them focus on the modeling for a single type of uncertainty in DSSE. Moreover, the direct impacts of uncertain DG outputs on DSSE are not fully addressed in unbalanced distribution systems. The existing ISE methods (*e.g.*, [13] and [18]) are established on the strong assumptions that the probability information or real-time measurements of DG outputs are available. In addition, the limitations of the solution strategies for the existing ISE models, including conservative estimation results and time-consuming solving process, persist.

In this paper, we propose a novel and fast ISE algorithm considering multiple uncertainties of DG outputs and line parameters in unbalanced distribution systems. As a solid reference to DSOs, the upper and lower bounds of state variables are provided by the proposed algorithm with hybrid SCADA and PMU measurements. First, based on the interval prediction for DG power outputs, an ISE model is formulated in interval arithmetic. Moreover, the model consider the line parameter uncertainty, and a weighted least square (WLS) criterion is integrated to deal with these hybrid measurements. Finally, a modified Krawczyk-operator (MKO) algorithm, which enables fast and accurate computation towards the bounds of state variables, is presented to obtain the interval solution of the proposed ISE model.

Main contributions of this paper are threefold.

- A general ISE framework is proposed to simultaneously formulate multiple uncertainties including imprecise line parameters, measurements with noises, and uncertain DG outputs. This framework is not limited to the assumptions that measurements or statistical information of DG outputs and accurate network parameters are available.
- This framework realizes the mutual transformation between different combinations of multiple uncertainties in unbalanced distribution systems.
- Algorithmically, the proposed algorithm solves the tight boundaries of state variables very efficiently, and the accuracy and efficiency are verified in comparison with thousands of times of MC simulations. Moreover, the algorithm is robust and not largely affected by the extent of the deviation of interval inputs relative to their true values.

## II. Hybrid DSSE Algorithm and Interval Arithmetic

### A. DSSE with Hybrid Measurements

In classical state estimators [2], the relationship between redundant measurements and state variables is depicted as:

$$\boldsymbol{z} = \boldsymbol{h}(\boldsymbol{x}) + \boldsymbol{e} \tag{1}$$

where $\boldsymbol{x}$ is an $n$-dimension state vector, and $\boldsymbol{z}$ is an $m$-dimension measurement vector; $\boldsymbol{h}(\boldsymbol{x})$ is the measurement function about $\boldsymbol{x}$; the measurement noise vector $\boldsymbol{e}$ obeys a Gaussian distribution $\boldsymbol{e}\sim N(0, \boldsymbol{R})$, where $\boldsymbol{R}$ is a covariance matrix and is usually considered diagonal (for instance, see [9] and [20]), $\boldsymbol{R} = diag[\sigma_1^2, \sigma_2^2, \dots, \sigma_m^2]$, and $\sigma_i^2$ denotes the variance of the $\boldsymbol{i}$th measurement error, $i = 1, 2, \dots, m$.

The state variables are obtained via a WLS criterion that minimizes the sum of weighted measurement residuals $J$ as:

$$\boldsymbol{x} = \arg\min\ J\ = \arg\min\ [\boldsymbol{z} - \boldsymbol{h}(\boldsymbol{x})]^T \boldsymbol{W} [\boldsymbol{z} - \boldsymbol{h}(\boldsymbol{x})] \tag{2}$$

where $\boldsymbol{W}$ is a weight matrix of measurements to quantify the trust levels of diverse measurements, and $\boldsymbol{W} = \boldsymbol{R}^{-1}$.

Optimal estimated states are solved iteratively by the Gauss-Newton method until each component of the vector $\Delta\boldsymbol{x}$ at each iteration is sufficiently small:

$$\partial \boldsymbol{J} / \partial \boldsymbol{x} = \boldsymbol{H}(\boldsymbol{x})^T \boldsymbol{W} [\boldsymbol{z} - \boldsymbol{h}(\boldsymbol{x})] = \boldsymbol{0} \tag{3}$$

$$\boldsymbol{H}(\boldsymbol{x})^T \boldsymbol{W} \boldsymbol{H}(\boldsymbol{x}) \Delta\boldsymbol{x} = \boldsymbol{H}(\boldsymbol{x})^T \boldsymbol{W} [\boldsymbol{z} - \boldsymbol{h}(\boldsymbol{x})] \tag{4}$$

where $\boldsymbol{H}(\boldsymbol{x})$ is the Jacobian matrix of the measurement function with respect to $\boldsymbol{x}$, and $\boldsymbol{H}(\boldsymbol{x}) = \partial \boldsymbol{h}(\boldsymbol{x}) / \partial \boldsymbol{x}$.

Further, recent research interests focus on the applications of PMUs to DSSE, since PMUs measure voltage and current phasors with high sampling precision and short update cycles [8]. Considering that a limited number of PMUs are installed in distribution systems due to their high technical and financial costs, hybrid state estimators incorporate conventional SCADA data with PMU data to improve estimation accuracy. Moreover, by adopting the state variables in the rectangular form, hybrid DSSE integrating PMU data results in a linear estimator, while conventional estimators with SCADA data are nonlinear [1].

In this paper, the hybrid estimator in [20] is used owing to its improved and recognized performance based on [2] and [19], where the voltage at a slack node (*i.e.*, a substation) and branch currents are chosen as state variables. In three-phase distribution systems, the state vector is expressed in rectangular coordinates as

$$\boldsymbol{x} = [v_{slack,r}^{a,b,c}, v_{slack,x}^{a,b,c}, i_{1r}^{a,b,c}, \cdots, i_{Nr}^{a,b,c}, i_{1x}^{a,b,c}, \cdots, i_{Nx}^{a,b,c}] \tag{5}$$

where $v_{slack,r}^{a,b,c}$ and $v_{slack,x}^{a,b,c}$ are the real and imaginary parts of the three-phase substation voltage, and $i_{lr}^{a,b,c}$ and $i_{lx}^{a,b,c}$ are the real and imaginary parts of the three-phase current at branch $l$, $l = 1, 2, \dots, N$. The superscripts $a$, $b$, and $c$ denote the phase indices.

In the hybrid estimator, power flows and power injections at loads are obtained by SCADA systems or pseudo-measurements, while PMUs provide the magnitude and phase angle measurements of voltages and currents. Moreover, the three-phase power measurements are converted into equivalent currents by

$$z_{I_{kr}} + j z_{I_{kx}} = \left[\frac{z_{P_k} + j z_{Q_k}}{V_k}\right]^* \tag{6.a}$$

$$z_{I_{pr}} + j z_{I_{px}} = \left[\frac{z_{P_p} + j z_{Q_p}}{V_p}\right]^* \tag{6.b}$$

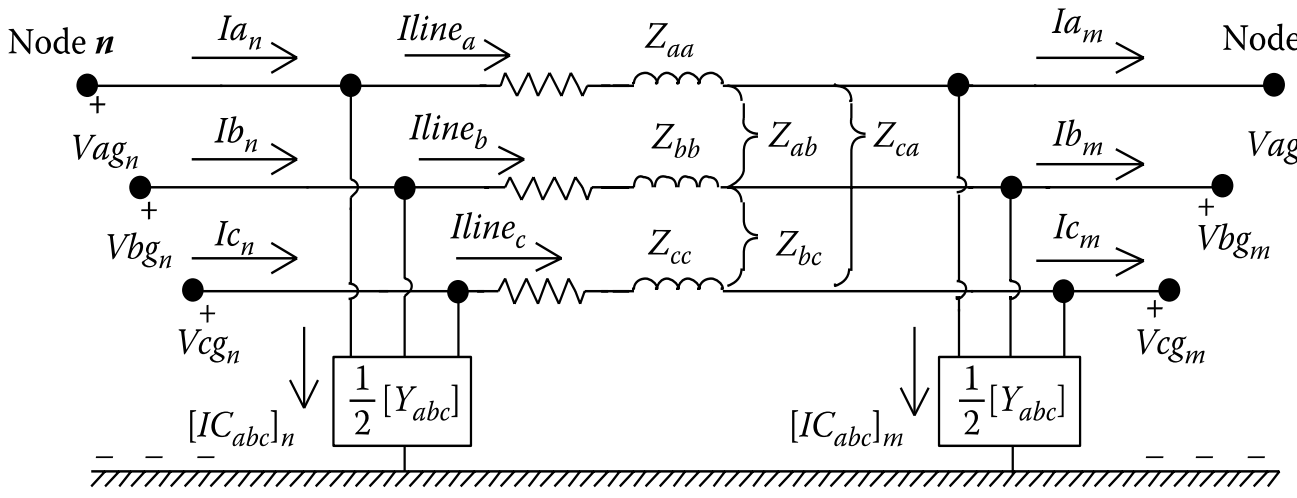

Fig.1. Three-phase line model in unbalanced distribution systems

where $z_{I_{kr}}$ and $z_{I_{kx}}$ (or $z_{I_{pr}}$ and $z_{I_{px}}$) are the real and imaginary parts of the current at node $k$ (or at branch $p$), and $V_k$ and $V_p$ are the voltage phasors at node $k$ and connected to branch $p$, repectively; $z_{P_k}$ and $z_{Q_k}$ (or $z_{P_p}$ and $z_{Q_p}$) denote the active and reactive powers at node $k$ (or at branch $p$). $[\cdot]^*$ denotes the complex conjugate.

The hybrid DSSE process in an unbalanced distribution system is iteratively implemented in the following steps [19]:

1) *Backward Sweep*: Get initial values of branch currents by a backward approach. An initial voltage at each node is set as the substation voltage $V_{slack}$, and (6.a) is used to calculate current injections through nodal power injections as follows:

$$z_{I_{kr}} + jz_{I_{kx}} = \left[\frac{z_{P_k} + jz_{Q_k}}{V_{slack}}\right]^* \quad (7)$$

Next, these injections are used to obtain branch currents.

2) *Forward Sweep*: The branch currents in step 1) and the substation voltage are used to calculate initial nodal voltages.

3) Calculate $\boldsymbol{h}(\boldsymbol{x})$, and then update system state variables as

$$\Delta\boldsymbol{x}^k = (\boldsymbol{H}(\boldsymbol{x}^k)^T\boldsymbol{W}\boldsymbol{H}(\boldsymbol{x}^k))^{-1}\boldsymbol{H}(\boldsymbol{x}^k)^T\boldsymbol{W}[\boldsymbol{z} - \boldsymbol{h}(\boldsymbol{x}^k)] \quad (8)$$

4) Update the branch currents by $\boldsymbol{x}^{k+1} = \boldsymbol{x}^k + \Delta\boldsymbol{x}^k$, then calculate nodal voltages by forward sweep in step 2).

5) If $\Delta\boldsymbol{x}^k$ is less than a pre-set tolerance, stop the iterative process. Otherwise, use these updated voltages to calculate the equivalent currents by (6), then go to step 3).

To improve the computational efficiency in the above iterative process, a linear approximation technique in [21] is used, where each nodal voltage is fixed as $V_{slack}$ to calculate equivalent currents in (6). This approximation is based on two observations for practical distribution systems, that 1) the voltage drops along feeders do not exceed 5%, since voltage phase differences are very small [22], and 2) nodal voltages are usually kept within normal operation limits (0.95 to 1.05 p.u.) [23]. Then, the Jacobian matrix is independent of $\boldsymbol{x}$ and highly sparse, and the general formulas (3) and (4) are updated with $\boldsymbol{h}(\boldsymbol{x}) = \boldsymbol{H}\boldsymbol{x}$ as:

$$\boldsymbol{\partial J}/\boldsymbol{\partial x} = \boldsymbol{H}^T\boldsymbol{W}\boldsymbol{z} - \boldsymbol{H}^T\boldsymbol{W}\boldsymbol{H}\boldsymbol{x} = 0 \quad (9)$$

$$\boldsymbol{x} = (\boldsymbol{H}^T\boldsymbol{W}\boldsymbol{H})^{-1}\boldsymbol{H}^T\boldsymbol{W}\boldsymbol{z} \quad (10)$$

The constant Jacobian elements of this estimator are briefly listed below, and more details about $\boldsymbol{H}$ can be referred to [20].

1) PMU voltage measurements

For the PMU voltage at node $k$, the measurement function is shown as

$$h(x) = h_{V_{kr}}^{a,b,c} + jh_{V_{kx}}^{a,b,c}$$
$$= (v_{slack,r}^{a,b,c} + jv_{slack,x}^{a,b,c}) - \sum\nolimits_{p\in\Im_k}(i_{pr}^{a,b,c} + ji_{px}^{a,b,c})z_p \quad (11)$$

where $\Im_k$ denotes a set of line segments from the slack node to node $k$, and $p \in \Im_k$ ; $z_p$ is the 3×3 impedance matrix of branch $p$, and $v_{slack,r}^{a,b,c}$, $v_{slack,x}^{a,b,c}$, $i_{pr}^{a,b,c}$, and $i_{px}^{a,b,c}$ as state variables are defined in (5). Also, the off-diagonal elements with non-zero values in $z_p$ reflect the coupling effect among three-phase lines, and $z_p$ can be shown as:

$$z_p = \begin{bmatrix} Z_{aa} & Z_{ab} & Z_{ac} \\ Z_{ba} & Z_{bb} & Z_{bc} \\ Z_{ca} & Z_{cb} & Z_{cc} \end{bmatrix}$$
$$= \begin{bmatrix} r_p^{aa} & r_p^{ab} & r_p^{ac} \\ r_p^{ba} & r_p^{bb} & r_p^{bc} \\ r_p^{ca} & r_p^{cb} & r_p^{cc} \end{bmatrix} + j\begin{bmatrix} x_p^{aa} & x_p^{ab} & x_p^{ac} \\ x_p^{ba} & x_p^{bb} & x_p^{bc} \\ x_p^{ca} & x_p^{cb} & x_p^{cc} \end{bmatrix}$$

where the diagonal and off-diagonal elements such as $Z_{aa}$ and $Z_{ab}$ denote the self-impedances and mutual impedances between two phases, respectively, shown as Fig.1. Besides, the mutual impedances between any two phases may not be the same each other due to the unbalanced nature of the system.

Take the A-phase voltage at node $k$ as an example, the non-zero Jacobian elements of the function (11) are expressed as:

$$\frac{\partial h_{V_{kr}}^a}{\partial v_{slack,r}^a} = 1 \qquad \frac{\partial h_{V_{kx}}^a}{\partial v_{slack,x}^a} = 1$$

$$\frac{\partial h_{V_{kr}}^a}{\partial i_{pr}^a} = -r_p^{aa} \quad \frac{\partial h_{V_{kr}}^a}{\partial i_{px}^a} = x_p^{aa} \quad \frac{\partial h_{V_{kx}}^a}{\partial i_{pr}^a} = -x_p^{aa} \quad \frac{\partial h_{V_{kx}}^a}{\partial i_{px}^a} = -r_p^{aa}$$

$$\frac{\partial h_{V_{kr}}^a}{\partial i_{pr}^b} = -r_p^{ab} \quad \frac{\partial h_{V_{kr}}^a}{\partial i_{px}^b} = x_p^{ab} \quad \frac{\partial h_{V_{kx}}^a}{\partial i_{pr}^b} = -x_p^{ab} \quad \frac{\partial h_{V_{kx}}^a}{\partial i_{px}^b} = -r_p^{ab}$$

$$\frac{\partial h_{V_{kr}}^a}{\partial i_{pr}^c} = -r_p^{ac} \quad \frac{\partial h_{V_{kr}}^a}{\partial i_{px}^c} = x_p^{ac} \quad \frac{\partial h_{V_{kx}}^a}{\partial i_{pr}^c} = -x_p^{ac} \quad \frac{\partial h_{V_{kx}}^a}{\partial i_{px}^c} = -r_p^{ac}$$

where $r_p^{a\varphi}$ denotes the mutual or self-resistance at branch $p$, $p \in \Im_k$, and $x_p^{a\varphi}$ denotes the mutual or self-reactance, $\varphi = a$, $b$, $c$. These phase indices are omitted for simplicity below.

2) PMU current measurements

For the PMU current at branch $p$, $h(x) = h_{I_{pr}} + jh_{I_{px}} = i_{pr} + ji_{px}$, and the only nonzero Jacobian elements are shown as:

$$\frac{\partial h_{I_{pr}}}{\partial i_{lr}} = \begin{cases} 1, & when\ p = l \\ 0, & elsewhere \end{cases} \qquad \frac{\partial h_{I_{px}}}{\partial i_{lx}} = \begin{cases} 1, & when\ p = l \\ 0, & elsewhere \end{cases}$$

where $l$ denotes the index of all branches.

3) Power measurements including line flows and power injections

For the power injections at node $k$, the Jacobian elements only has nonzero values of +1 and −1, since the measurement function holds based on Kirchhoff's current law:

$$h(x) = h_{I_{kr}} + jh_{I_{kx}} = i_{in,r} + ji_{in,x} - \sum(i_{out,r} + ji_{out,x}) \quad (12)$$

where $i_{in,r} + ji_{in,x}$ and $i_{out,r} + ji_{out,x}$ as state variables denote the input and output currents at node $k$, respectively.

Besides, the Jacobian elements of the equivalent currents at branch $p$ from the corresponding line flows only have nonzero values of +1, similar to PMU current measurements.

### B. *Interval Arithmetic and Interval Prediction of DG*

An interval number is defined as a compact set $[a] = [a_l, a_u] = \{a \in \mathbb{R}\,|a_l \le a \le a_u\}$, and similarly, an interval vector is defined as a vector with interval elements [15].

When meters are not available at DG locations, effective forecasting techniques are utilized to obtain DG power outputs as pseudo-measurements for achieving system observability [1]. Due to the difficulty in accurate forecasts for instantaneous wind speeds or solar radiations, their forecast errors inevitably result in considerable forecast errors of DG outputs [24]. Hence, DG outputs in the interval form are modeled to quantify the uncertainty levels in interval predictions, which is more feasible in practice [15]. Conventional pseudo-measurements originate from the historical or forecast data on generator production and load consumption. Moreover, they obey Gaussian distributions with high-level noises as in [7] or are represented as other known distribution information as in [8]. In the proposed algorithm, we relax these assumptions and use the interval prediction of DG outputs. Further, the interval DG outputs are deemed as another form of pseudo-measurements in DSSE to obtain the interval estimate of states in the subsequent section.

## III. General ISE Framework and Proposed Algorithm

### A. *ISE Model with Multiple Uncertainties*

In this section, an ISE model with multiple uncertainties in DSSE is proposed, where measurements with noise, uncertain DG outputs, and imprecise line parameters are considered. The above highly efficient estimator is used to achieve fast monitoring of distribution networks with these uncertainties.

The impacts of the uncertainty sources on the deterministic estimation model (9) are analyzed and then updated into an interval estimation model as

$$\begin{bmatrix}[\boldsymbol{H}_1]\\ \boldsymbol{H}_2\end{bmatrix}^T \begin{bmatrix}\boldsymbol{W}_1 & 0\\ 0 & \boldsymbol{I}\end{bmatrix} \begin{bmatrix}[\boldsymbol{H}_1]\\ \boldsymbol{H}_2\end{bmatrix}[\boldsymbol{x}] = \begin{bmatrix}[\boldsymbol{H}_1]\\ \boldsymbol{H}_2\end{bmatrix}^T \begin{bmatrix}\boldsymbol{W}_1 & 0\\ 0 & \boldsymbol{I}\end{bmatrix} \begin{bmatrix}[\boldsymbol{z}_1]\\ [\boldsymbol{z}_2]\end{bmatrix} \tag{13}$$

where $[\boldsymbol{x}]$ is an interval state vector, and $[\boldsymbol{x}] \in \mathbb{R}^{n\times 1}$; $[\boldsymbol{z}_1]$ denotes an interval measurement vector, and $[\boldsymbol{z}_1] \in \mathbb{R}^{m_1\times 1}$, while $[\boldsymbol{z}_2]$ denotes an interval vector of DG power outputs, and $[\boldsymbol{z}_2] \in \mathbb{R}^{m_2\times 1}$. $[\boldsymbol{H}_1]$ and $\boldsymbol{H}_2$ are the Jacobian matrices of the measurements and the DG outputs, and $[\boldsymbol{H}_1] \in \mathbb{R}^{m_1\times n}$, $\boldsymbol{H}_2 \in \mathbb{R}^{m_2\times n}$; $\boldsymbol{W}_1$ is the weighted matrix of the measurements, and $\boldsymbol{W}_1 \in \mathbb{R}^{m_1\times m_1}$; $\boldsymbol{I}$ is an identity matrix, and $\boldsymbol{I} \in \mathbb{R}^{m_2\times m_2}$.

The top-row equation describes the relationship between the measurements and $[\boldsymbol{x}]$; the bottom-row equation provides the constraints for the states related to DG outputs. The uncertain outputs of DG are modeled as pseudo-measurements of the system according to their interval predictions mentioned in Section II-B. Note that only solving the equation set at the top row in (13) may not obtain the solutions of these state variables. This is owing to practical engineering concerns that the measurements of DG outputs are not available at all DG locations. As a result, the top-row system with measurements may be unobservable by DSSE due to the lack of necessary measurements [6]. The formula (13) ensures that an interval solution not only meets a WLS criterion for all measurements but also follows the relationship with these interval DG outputs.

According to different uncertainty sources, the details of (13) are individually discussed for clarity.

1) *Measurements.* Power measurements are converted to equivalent currents by (7), then $[\boldsymbol{z}_1]$ is expressed as

$$[\boldsymbol{z}_1] = \begin{bmatrix}[\boldsymbol{U}_l, \boldsymbol{U}_u]\\ [\boldsymbol{I}_l, \boldsymbol{I}_u]\\ [\boldsymbol{I}_{eq,l}, \boldsymbol{I}_{eq,u}]\end{bmatrix} \tag{14}$$

where $\boldsymbol{U}_l$ and $\boldsymbol{I}_l$ represent the lower bounds of voltage and current vectors from PMU measurements, and $\boldsymbol{I}_{eq,l}$ is the lower bound of equivalent current measurements, while $\boldsymbol{U}_u$, $\boldsymbol{I}_u$, and $\boldsymbol{I}_{eq,u}$ represent the corresponding upper bounds.

The $3\sigma$ deviation criterion about the mean in a Gaussian distribution covers more than 99.7% of the area of the distribution is used to obtain $[\boldsymbol{z}_1]$ based on measurements with noises [7]. For any measurement $Z_m$ with a random noise $\boldsymbol{e}\sim N(0,\sigma^2)$, $Z_m \in [Z_0 - 3\sigma, Z_0 + 3\sigma]$, where $Z_0$ represents the true value. Hence, $Z_0 \in [Z_m - 3\sigma, Z_m + 3\sigma]$, and this measurement is modeled as an interval enclosing the corresponding true value.

2) *DG outputs*. The upper and lower bounds of DG outputs as pseudo-measurements are transformed to equivalent currents by (7), respectively. $[\boldsymbol{z}_2] = [\boldsymbol{I}_{DG,l}, \boldsymbol{I}_{DG,u}]$, where $\boldsymbol{I}_{DG,l}$ and $\boldsymbol{I}_{DG,u}$ represent the lower and upper bounds of the equivalent currents. As discussed in (12), the Jacobian matrix $\boldsymbol{H}_2$ related with DG outputs only has nonzero elements of +1 and $-1$. Besides, we relax the strong assumption that the statistical information of these DG outputs is known, *i.e.*, there is no requirement of the knowledge of the mean and covariance of DG outputs. Such correlations can be translated into respective DG output intervals and then the proposed method can be still applicable. The detailed consideration of the DG correlation will be left for our future work.

3) *Line Parameters.* The uncertainty of line parameters is evaluated in a range based on their nameplate values [17]. In the adopted estimator, the uncertain line parameters are only present at the locations corresponding to PMU voltage measurements, *i.e.*, in (11) and in the matrix $[\boldsymbol{H}_1]$ of (13). Hence, with the line parameter uncertainty, $\boldsymbol{H}_1$ is updated into an interval matrix $[\boldsymbol{H}_1]$ in (13).

### B. *General ISE Framework*

The impacts of multiple uncertainties on ISE are decoupled in (13), and thus an ISE framework is proposed to deal with different combinations of multiple uncertainties to obtain the bounds of state variables. The model (13) is expressed in a compact form:

$$[\boldsymbol{H}]^T \boldsymbol{W} [\boldsymbol{H}][\boldsymbol{x}] = [\boldsymbol{H}]^T \boldsymbol{W} [\boldsymbol{z}] \tag{15}$$

where $[\boldsymbol{H}] = \begin{bmatrix}[\boldsymbol{H}_1]\\ \boldsymbol{H}_2\end{bmatrix}$, and $[\boldsymbol{H}] \in \mathbb{R}^{m\times n}$; $\boldsymbol{W} = \begin{bmatrix}\boldsymbol{W}_1 & \boldsymbol{0}\\ \boldsymbol{0} & \boldsymbol{I}\end{bmatrix}$, and $\boldsymbol{W} \in \mathbb{R}^{m\times m}$; $[\boldsymbol{z}] = \begin{bmatrix}[\boldsymbol{z}_1]\\ [\boldsymbol{z}_2]\end{bmatrix}$, and $[\boldsymbol{z}] \in \mathbb{R}^{m\times 1}$. $m$ is the total number of the measurements and DG prediction intervals, and $m = m_1 + m_2$.

In order to avoid computing the interval multiplication in $[\boldsymbol{H}]^T \boldsymbol{W} [\boldsymbol{H}]$, which is computationally expensive, a dummy interval vector $[\boldsymbol{y}]$ is introduced into (15) as suggested in [25]. Then, an equivalent equation is obtained:

$$\begin{bmatrix}[\boldsymbol{H}] & -\boldsymbol{I}\\ 0 & [\boldsymbol{H}]^T\boldsymbol{W}\end{bmatrix}\begin{bmatrix}[\boldsymbol{x}]\\ [\boldsymbol{y}]\end{bmatrix} = \begin{bmatrix}[\boldsymbol{z}]\\ 0\end{bmatrix} \tag{16}$$

where $\boldsymbol{I}$ is an identity matrix, and $\boldsymbol{I} \in \mathbb{R}^{m\times m}$; $[\boldsymbol{y}] \in \mathbb{R}^{m\times 1}$.

The formula (16) is further expressed below for brevity:

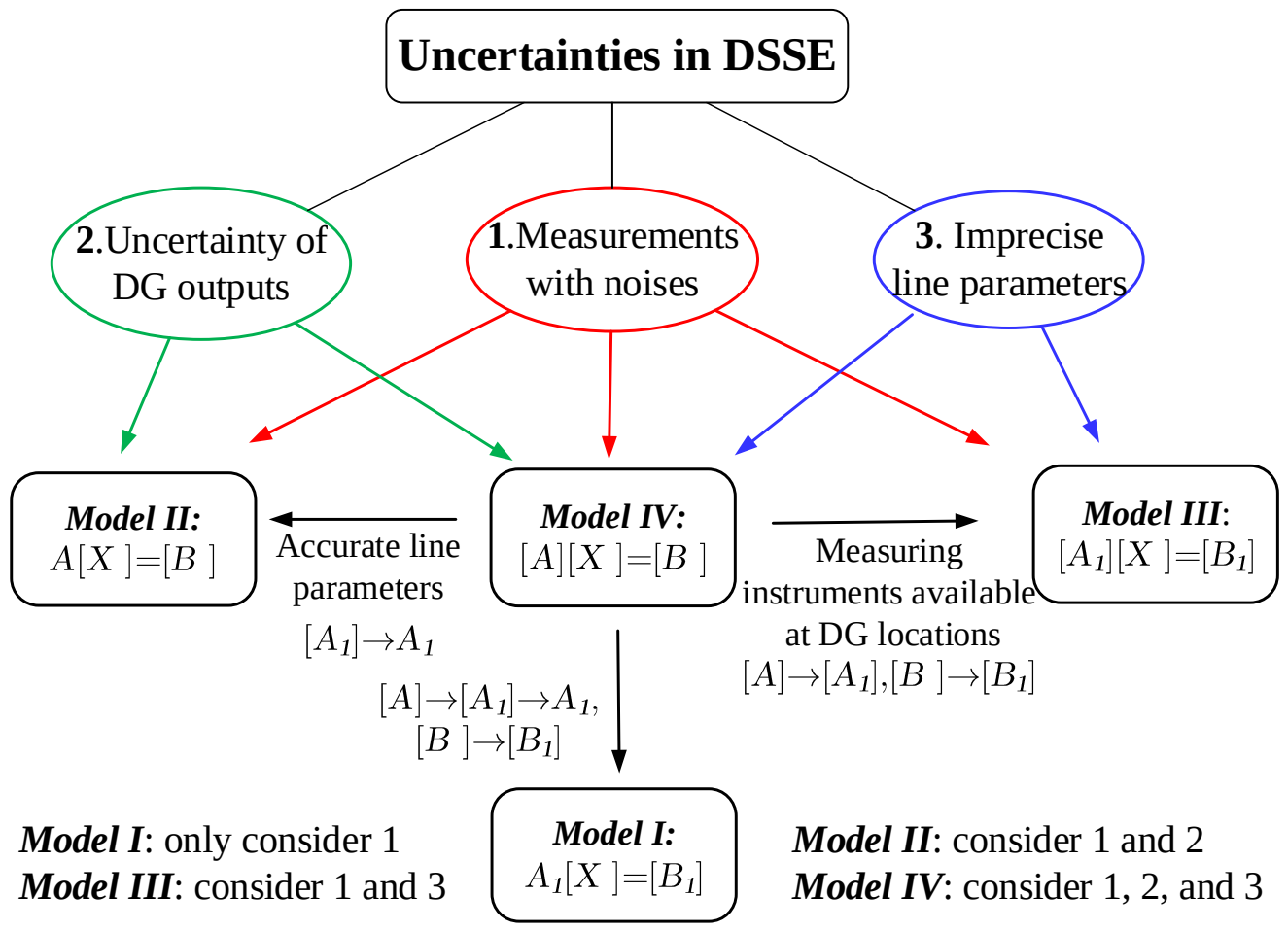

Fig. 2. Uncertainties of DSSE in interval arithmetic. The model complexity increases from *Model I* to *Model IV*.

$$[\boldsymbol{A}][\boldsymbol{X}] = [\boldsymbol{B}] \tag{17}$$

where $[\boldsymbol{A}] = \begin{bmatrix} [\boldsymbol{H}] & -\boldsymbol{I} \\ 0 & [\boldsymbol{H}]^T\boldsymbol{W} \end{bmatrix}$, and $[\boldsymbol{A}] \in \mathbb{R}^{(m+n)\times(m+n)}$; $[\boldsymbol{B}] = \begin{bmatrix} [\boldsymbol{z}] \\ 0 \end{bmatrix}$, and $[\boldsymbol{B}] \in \mathbb{R}^{(m+n)\times 1}$; $[\boldsymbol{X}] = \begin{bmatrix} [\boldsymbol{x}] \\ [\boldsymbol{y}] \end{bmatrix}$, and $[\boldsymbol{X}] \in \mathbb{R}^{(m+n)\times 1}$.

The model (17) is straightforward to realize the mutual transformation between different combinations of uncertainties. This transformation is shown in Fig.2. A general ISE model considering three types of uncertainties is formulated as (17) and named *Model IV*, while *Model I* is a basic ISE model only considering the measurements with noises. When meters are available at all DG units, with the DG measurements merged into $[\boldsymbol{z}_1]$, *Model IV* is simplified into *Model III*, where $[\boldsymbol{A}_1] = \begin{bmatrix} [\boldsymbol{H}_1] & -\boldsymbol{I} \\ 0 & [\boldsymbol{H}_1]^T\boldsymbol{W}_1 \end{bmatrix}$, and $[\boldsymbol{B}_1] = \begin{bmatrix} [\boldsymbol{z}_1] \\ 0 \end{bmatrix}$. In another case, *Model IV* is simplified into *Model II*, when parameter identification techniques or the assumption with accurate line parameters are adopted, *i.e.*, $[\boldsymbol{H}_1]$ is fixed as $\boldsymbol{H}_1$.

### *C. MKO Algorithm for Solving ISE Models*

In this section, an MKO algorithm in interval arithmetic is proposed to efficiently solve the compact ISE model in (17). The interval symbols [ ] are omitted here for clarity. For the interval system $\boldsymbol{AX} = \boldsymbol{B}$, its solution set in the interval form is expressed as $\sum(\boldsymbol{A}, \boldsymbol{B}) = \{\widetilde{\boldsymbol{X}}\text{: } \boldsymbol{a}\widetilde{\boldsymbol{X}} = \boldsymbol{b},\ \forall \boldsymbol{a} \in \boldsymbol{A} \text{ and } \forall \boldsymbol{b} \in \boldsymbol{B}\}$, where $\widetilde{\boldsymbol{X}}$, $\boldsymbol{a}$, and $\boldsymbol{b}$ are the corresponding deterministic vectors or matrices. Moreover, its interval solution hull $\boldsymbol{X}$ is the interval vector with the smallest radius containing $\sum(\boldsymbol{A}, \boldsymbol{B})$.

A Krawczyk operator is widely employed as a solver for linear interval equations [26]. The core of this operator is to utilize an approximate interval solution $\boldsymbol{X}^{(0)}$ that contains the final solution hull as an initial value, then gradually approach the final solution hull by the following iterative process:

$$\boldsymbol{X}^{(i+1)} = (\boldsymbol{CB} + (\boldsymbol{I} - \boldsymbol{CA})\boldsymbol{X}^{(i)}) \cap \boldsymbol{X}^{(i)} \tag{18}$$

where $\boldsymbol{C}$ is a preconditioning point matrix, and the inverse of $\boldsymbol{C}$ is the midpoint matrix of $\boldsymbol{A}$, expressed as

$$\boldsymbol{C}^{-1} = Mid[\boldsymbol{A}] = \begin{bmatrix} Mid[\boldsymbol{a}_{11}] & \cdots & Mid[\boldsymbol{a}_{1,m+n}] \\ \vdots & \ddots & \vdots \\ Mid[\boldsymbol{a}_{m+n,1}] & \cdots & Mid[\boldsymbol{a}_{m+n,m+n}] \end{bmatrix}$$

where $Mid[\cdot]$ is the median function of an interval variable.

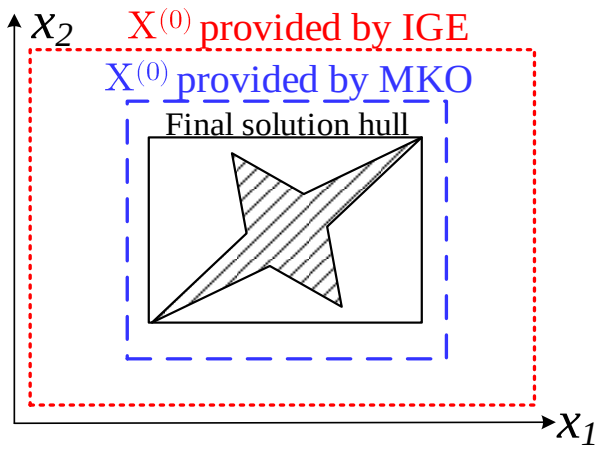

Fig. 3. Solution hulls of a 2-dimensional interval equation

It is deduced that with this matrix $\boldsymbol{C}$ that satisfies $\|\boldsymbol{I} - \boldsymbol{CA}\| < 1$, where $\|\cdot\|$ is any subordinate norm, (18) converges according to the fixed point theorem [27]. The iterative process runs until $\|\boldsymbol{X}^{(i+1)} - \boldsymbol{X}^{(i)}\|_\infty \le \varepsilon$, and we set $\varepsilon = 10^{-4}$. Hence, the Krawczyk operator avoids the issue of interval extension when the tolerance of variables in the iterative process is sufficiently small, and the interested readers can refer to the proof in [27].

Next, the above Krawczyk operator is modified to improve algorithmic performance in both accuracy and efficiency. In the proposed MKO algorithm, two computational strategies, Strategy One and Strategy Two, are jointly used to solve (17) quickly and accurately:

1) **Strategy One**: Start from an initial solution $\boldsymbol{X}^{(0)}$, which is closer to the final solution hull compared to the one that IGE produces.

In the Krawczyk operator, an initial solution $\boldsymbol{X}^{(0)}$, which meets $\widetilde{\boldsymbol{X}} \in \boldsymbol{X}^{(0)}$ and $\sum(\boldsymbol{A}, \boldsymbol{B}) \subseteq \boldsymbol{X}^{(0)}$, is required to start the iterative process. A straightforward approach to getting $\boldsymbol{X}^{(0)}$ is IGE, which is used in [18] as an extension of Gaussian elimination in interval arithmetic. However, IGE largely expands the widths of interval solutions due to its over-conservatism [28]. In addition, IGE is expensive to compute since its forward elimination and back substitution procedure cannot be parallelized.

To address the limitations of IGE, a tighter $\boldsymbol{X}^{(0)}$ is obtained by the following theorem [29].

**Theorem 1.** If $\boldsymbol{C}$ satisfies $\|\boldsymbol{I} - \boldsymbol{Ca}\| = \beta < 1$, $\widetilde{\boldsymbol{X}} = \boldsymbol{a}^{-1}\boldsymbol{b}$ and $\|\cdot\|$ is any subordinate norm, then $\|\widetilde{\boldsymbol{X}}\| \le \frac{\|\boldsymbol{Cb}\|}{1-\beta}$.

*Proof.* From $\boldsymbol{a}\widetilde{\boldsymbol{X}} = \boldsymbol{b}$, we have $\widetilde{\boldsymbol{X}} = \boldsymbol{Cb} + (\boldsymbol{I} - \boldsymbol{Ca})\widetilde{\boldsymbol{X}}$, and hence

$$\|\widetilde{\boldsymbol{X}}\| \le \|\boldsymbol{Cb}\| + \|\boldsymbol{I} - \boldsymbol{Ca}\| \, \|\widetilde{\boldsymbol{X}}\| \to (1-\beta)\|\widetilde{\boldsymbol{X}}\| \le \|\boldsymbol{Cb}\|$$

where $\beta < 1$ exists for $\boldsymbol{C}$, which is the inverse of the midpoint matrix of $\boldsymbol{A}$. ■

Since $\|\boldsymbol{Cb}\|_\infty \le \|\boldsymbol{CB}\|_\infty$ and $\|\boldsymbol{I} - \boldsymbol{Ca}\|_\infty \le \|\boldsymbol{I} - \boldsymbol{CA}\|_\infty$, an initial vector $\boldsymbol{X}^{(0)}$ is defined as

$$\boldsymbol{X}^{(0)} = ([-\alpha, \alpha], \dots, [-\alpha, \alpha])^T \tag{19}$$

where $\alpha = \frac{\|\boldsymbol{CB}\|_\infty}{1-\beta}$ and $\beta = \|\boldsymbol{I} - \boldsymbol{CA}\|_\infty$. $\|\cdot\|_\infty$ is the infinite norm of a vector.

The tighter initial solution in (19), which is closer to the final solution hull compared to the IGE solution, largely improves the efficiency of the iterative process. A schematic diagram in Fig.3 illustrates the phenomenon, where the accurate solution set of the interval equation constitutes a star-shaped area [26].

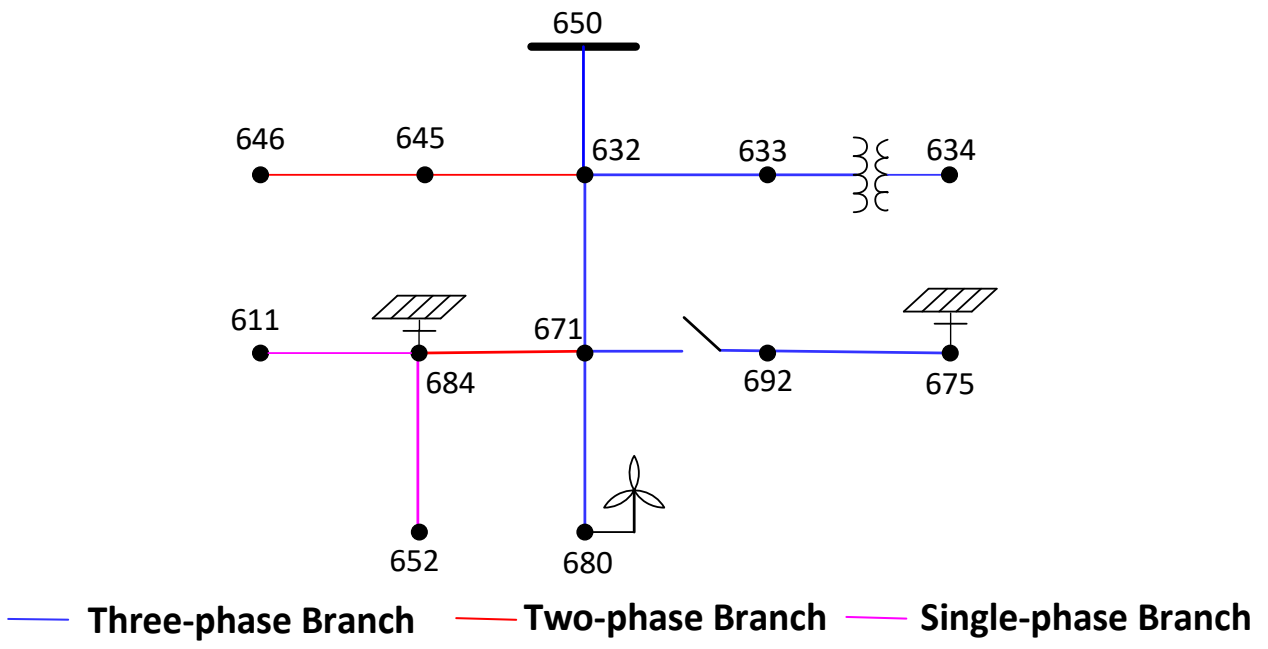

Fig. 4. One-line diagram of unbalanced 13-bus system

TABLE I
DG PLACEMENTS IN A 123-BUS SYSTEM

| # | No. node | Type | Phase | # | No. node | Type | Phase |
|---|---|---|---|---|---|---|---|
| 1 | 14 | PV | C | 4 | 250 | WTG | A, B, C |
| 2 | 61 | WTG | A, B, C | 5 | 300 | PV | A |
| 3 | 151 | WTG | A, B, C | 6 | 450 | PV | A |

2) **Strategy Two**: Modify an enclosure of the difference between the solution at the $i$ th iteration $\boldsymbol{X}^{(i)}$ and an approximate solution, rather than $\boldsymbol{X}^{(i)}$. This modification, combined with Strategy One, guarantees that the proposed algorithm produces the final solution at least as tight as the original Krawczyk operator.

First, the approximate solution $\boldsymbol{X}_s$, which is a point solution located at the center of the solution space, is calculated by multiplying $\boldsymbol{C}$ by the midpoint vector of $\boldsymbol{B}$:

$$\boldsymbol{X}_s = \boldsymbol{C}(Mid[\boldsymbol{B}]) \tag{20}$$

Let $\boldsymbol{d} = \boldsymbol{X} - \boldsymbol{X}_s$, and get an equivalent interval equation $\boldsymbol{Ad} = \boldsymbol{B} - \boldsymbol{AX}_s$. The initial solution for this equation is $\boldsymbol{d}^{(0)} = \boldsymbol{X}^{(0)} - \boldsymbol{X}_s$, where $\boldsymbol{X}^{(0)}$ is calculated by (19).

Applying (19) to the enclosure $\boldsymbol{d}^{(i+1)}$ gives a modified residual Krawczyk iterative process

$$\boldsymbol{d}^{(i+1)} = (\boldsymbol{C}(\boldsymbol{B} - \boldsymbol{AX}_s) + (\boldsymbol{I} - \boldsymbol{CA})\boldsymbol{d}^{(i)}) \cap \boldsymbol{d}^{(i)} \tag{21}$$

until $\|\boldsymbol{d}^{(i+1)} - \boldsymbol{d}^{(i)}\|_\infty \le \varepsilon$.

A final solution is computed by $\boldsymbol{X}^{(i+1)} = \boldsymbol{X}_s + \boldsymbol{d}^{(i+1)}$. The final solution produced by (21) is at least as tight as the original Krawczyk operator in (18), which is verified through the sub-distributive law for interval arithmetic.

**Theorem 2. Sub-distributive Law [28].** For interval variables $x$, $y$, and $z$, the law holds

$$x(y+z) \subseteq xy + xz \tag{22}$$

Apply Theorem 2 into (21), and we have

$$\boldsymbol{X}_s + \boldsymbol{C}(\boldsymbol{B} - \boldsymbol{AX}_s) + (\boldsymbol{I} - \boldsymbol{CA})\boldsymbol{d}^{(i)} \supseteq \boldsymbol{CB} + (\boldsymbol{I} - \boldsymbol{CA})(\boldsymbol{X}_s + \boldsymbol{d}^{(i)}) = \boldsymbol{CB} + (\boldsymbol{I} - \boldsymbol{CA})\boldsymbol{X}^{(i)} \tag{23}$$

where $\boldsymbol{X}^{(i)} = \boldsymbol{X}_s + \boldsymbol{d}^{(i)}$.

The formula (23) implies that $\boldsymbol{X}^{(i+1)} = \boldsymbol{X}_s + \boldsymbol{d}^{(i+1)} \supseteq (\boldsymbol{CB} + (\boldsymbol{I} - \boldsymbol{CA})\boldsymbol{X}^{(i+1)}) \cap \boldsymbol{X}^{(i+1)}$, *i.e.*, the final solution hull provided by the original Krawczyk operator contains the one that the MKO algorithm solves. Hence, the proposed algorithm obtains a tighter boundary than the original Krawczyk operator. Moreover, if $\boldsymbol{A}$ and $\boldsymbol{B}$ are thin (an interval with zero radius is defined as a thin interval), the residual $\boldsymbol{B} - \boldsymbol{AX}_s$ is enclosed with fewer rounding errors than $\boldsymbol{B}$ [29, Chapter 4]. These features lead to the higher accuracy and less memory space of the proposed algorithm, since many thin interval elements exist due to the highly sparse $\boldsymbol{A}$ and $\boldsymbol{B}$ shown in (16), i.e., $0 \equiv [0,0]$.

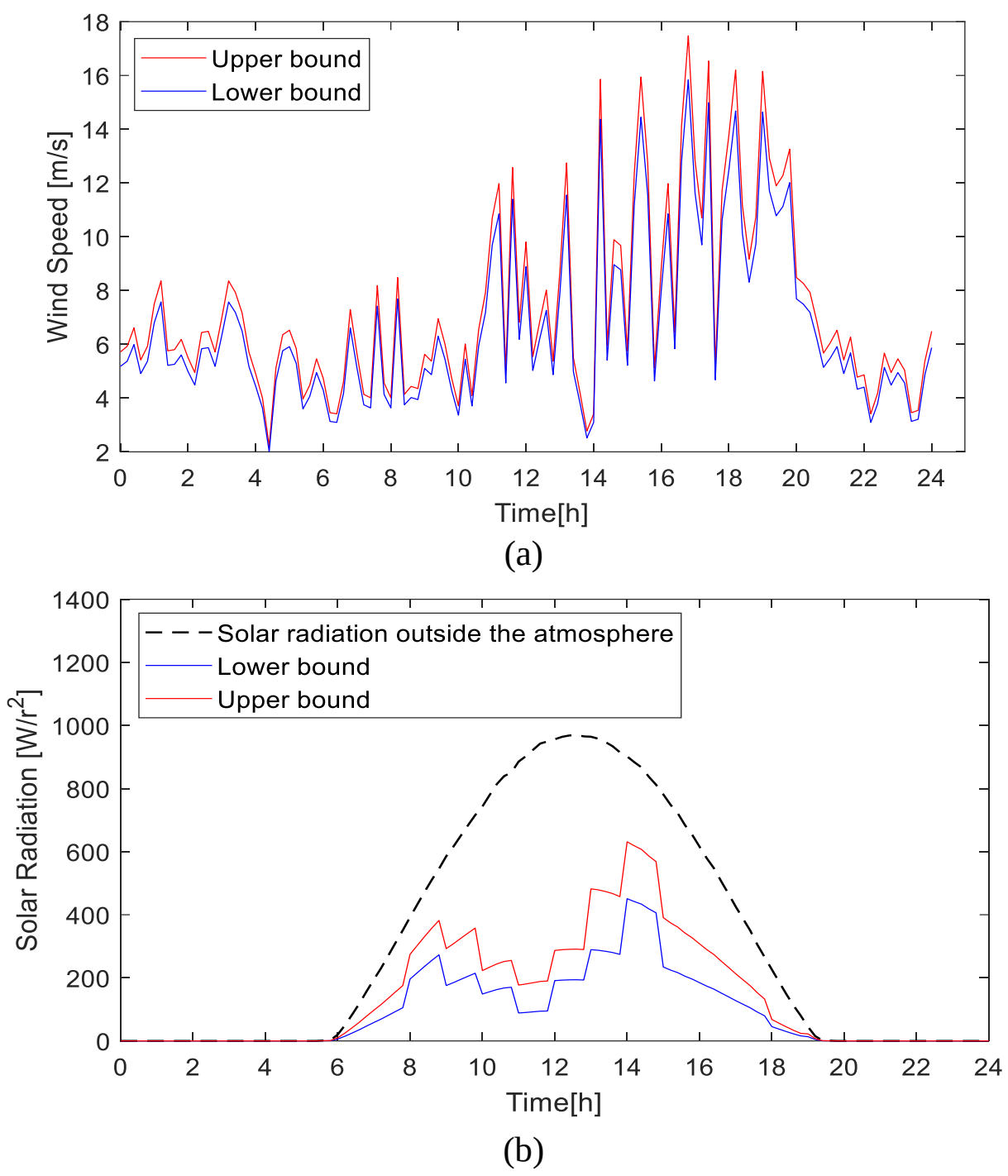

Fig. 5. DG profiles during one day (a) Wind speed (b) Solar radiation

## IV. SIMULATION RESULT

The proposed algorithm is tested on unbalanced IEEE 13-bus and 123-bus distribution systems [30]. The 13-bus system is modified by adding two single-phase PV units at buses 675 and 684, and a three-phase wind farm at bus 680, shown in Fig. 4. The 123-bus system is modified by adding six DG units, and the installation details of these DG units are listed in Table I.

Based on the weather data [31] shown in Fig. 5, at 12:00 am, the wind speed interval is [8.886, 9.805] m/s, and the solar radiation interval is [191.246, 286.870] W/m$^2$. The interval outputs of PVs and WTGs are obtained by the method in [15], and constant power factors are used [9]: [106.72, 149.53] kW for PVs, 0.95 lagging and [84.52, 103.31] kW for WTGs, 0.85 lagging. For these DGs, the upper and lower bounds of reactive power are calculated by $Q_{DG,u} = P_{DG,u}\tan\phi$ and $Q_{DG,l} = P_{DG,l}\tan\phi$, where $P_{DG,u}$ and $P_{DG,l}$ are the upper and lower bounds of power outputs, and $\phi$ is the power factor angle.

For simulation purposes, the deterministic DG outputs and constant line parameters that fall in the corresponding intervals are used to obtain true values of voltages, currents, and powers by the power flow program. Measurements are obtained by adding Gaussian noises to these true values. The following conditions are applied to maximum percentage errors of hybrid measurements in Table II: 0.7% for magnitudes and 0.7 crad ($10^{-2}$ rad) for phase angles in PMU data [32], 2% for active and reactive powers of SCADA data, and 10% for active and reactive powers at load nodes as pseudo-measurements [13]. Besides, these hybrid measurements with different sampling rates can be pre-processed for synchronization by the method in [33]. Note that meters and statistical information of power outputs at some DGs are not available, *e.g.*, DGs at buses 680 and 675 in the 13-bus system. Conventional DSSE defines such systems as unobservable, *i.e.*, DSSE fails due to lack of key measurements. In these cases, the proposed algorithm provides the ranges of state variables with the aid of prediction intervals of DG outputs via *Model II* and *Model IV*.

TABLE II
MEASUREMENT LOCATIONS IN TEST SYSTEMS

| Measurement Types | Placement Locations | |
|---|---|---|
| | 13-bus System | 123-bus System |
| SCADA | 632-633, 645-646, 684-652 | 1-7, 9-14, 15-16,13-52, 18-35, 44-45, 57-60, 76-77, 86-87, 99-100, 110-112 |
| PMU | 650, 671 | 149, 8, 25, 54, 97, 108 |

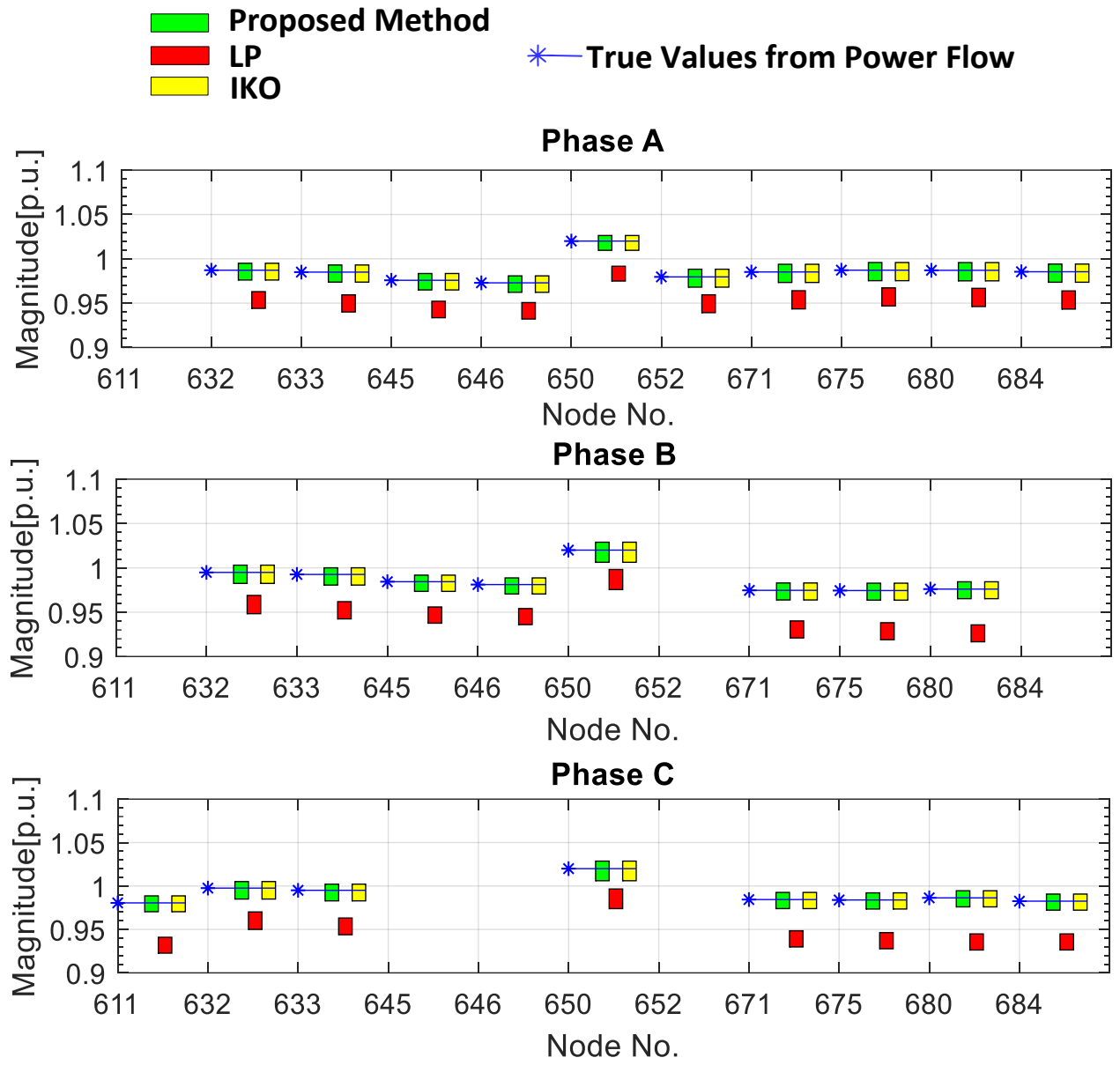

Fig. 6. Voltage magnitude results of the proposed algorithm

## A. Result Analysis

The proposed algorithm is tested with *Model II* on the 13-bus system, provided that accurate line parameters are known. In this section, we display the voltage magnitude results of different methods considering that they are more concerned by DSOs in practice, and these voltage magnitudes are calculated from state variables.

The estimation results of the proposed algorithm, a linear programming (LP)-based method, and the IKO method in [18] are compared with true values of voltages, and the three-phase voltage magnitudes in these methods are intuitively displayed in Fig.6. The proposed method provides the tight bounds that all possible state variables fall in under multiple uncertainties, and these ranges contain the true values of voltages. However, the voltages solved by the LP-based method deviate a lot from the true values at some buses and even exceed the normal operation voltage range (0.95 to 1.05 p.u.). This is because this method does not consider the various weights of hybrid measurements in the test system.

Moreover, the sums of these voltage widths between the proposed method and the IGE-based method are numerically shown in Table III. The widths of voltages for the proposed algorithm are narrower than the IGE-based method.

TABLE III
DETAILED COMPARISON IN VOLTAGE RESULTS

| Sum of Widths [V] | Proposed Method | | IKO | |
|---|---|---|---|---|
| | Re. Part | Im. Part | Re. Part | Im. Part |
| Phase A | 471.05 | 260.47 | 471.08 | 260.49 |
| Phase B | 365.64 | 214.67 | 365.65 | 214.69 |
| Phase C | 393.31 | 215.95 | 393.34 | 215.97 |

## B. Effect of Parameter Uncertainty

The proposed algorithm is applied to two situations in which the line parameters are determined or in certain ranges, via *Model II* and *Model IV*, respectively. To investigate the influence of uncertain line parameters, two cases are designed:

*Case 1*: Constant line parameter vector, $\boldsymbol{P}_0$.

*Case 2*: Line parameters change in $[0.95\boldsymbol{P}_0, 1.05\boldsymbol{P}_0]$.

Two indices $Q_1$ and $Q_2$ are used to evaluate the precision of interval estimation:

$$Q_1 = \frac{1}{n}\sum_{i=1}^{n}(\overline{x}_i - \underline{x}_i) \tag{25}$$

$$Q_2 = \max(|\overline{x}_i - x_i|, |x_i - \underline{x}_i|) \tag{26}$$

where $\overline{x}_i$ and $\underline{x}_i$ are the upper and lower bounds of the $i$th interval variable, and $x_i$ denotes the true value of the $i$th state variable. $Q_1$ is the average value of interval widths, and $Q_2$ is the maximum deviation relative to the true values. The interval estimation with smaller $Q_1$ and $Q_2$ has better accuracy.

To verify the effectiveness of the proposed algorithm, in Case 1 and Case 2, the deterministic DSSE algorithm in [20] without linear approximation runs for 3000 times of MC trials. In these MC trials, random DG outputs in their predication intervals are regarded as the inputs to the DSSE method in Case 1, while the random combinations of deterministic DG outputs and line parameters that fall in the corresponding intervals are used in Case 2. The minimum and maximum values of these state variables in all MC trials are collected and compared with the interval estimation results of the proposed algorithm.

Limited by space, the voltage results at the even-numbered nodes of the 123-bus system are depicted in Fig. 7, where the true values of these states are also marked. The interval results of both methods are shown as similar in the two cases. Concretely, the accuracy indices $Q_1$ and $Q_2$ in Fig. 7(b) are 0.0196 and 0.0163 in the proposed algorithm, while they are 0.0182 and 0.0171 in MC simulations. These results illustrate that under multiple uncertainties, the proposed algorithm obtains the tight boundaries of state variables via a single run, compared with thousands of times of MC runs. Also, the comparison between Fig. 7(a) and 7(b) demonstrates that the line parameter uncertainty further intensifies the variability of state variables. With these imprecise line parameters, the proposed method provides the ranges that all possible values of states fall in, as a reliable reference to system operators, shown as Fig. 7(b).

## C. Robustness Tests

The robustness of the proposed algorithm is tested on the 123-bus system. Based on Case 2, three cases are established, where the true values of DG outputs lie on the edge of interval predictions, *i.e.*, asymmetric intervals. These cases are designed below.

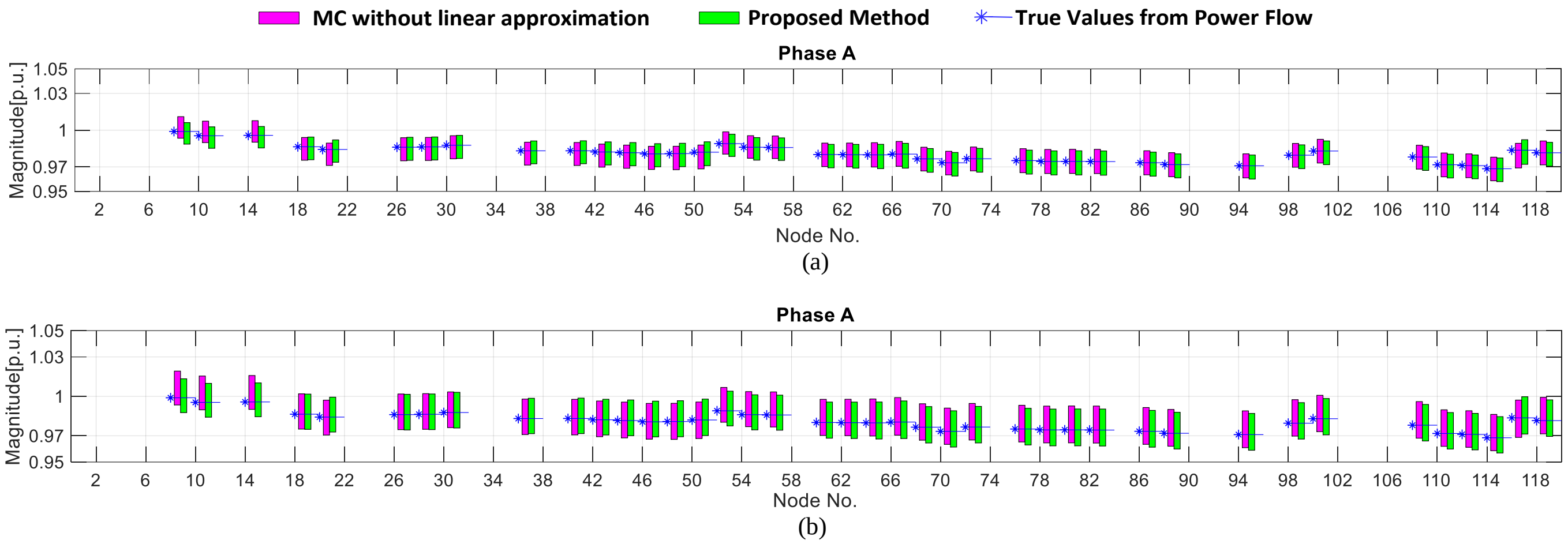

Fig. 7. Voltage magnitude results in the 123-bus system (a) Case 1: Without line uncertainty (b) Case 2: With line uncertainty. Under multiple uncertainties, although a linear approximation technique is adopted, the interval results of our algorithm are similar to the ones of MC simulations without this approximation.

TABLE IV
ESTIMATION ACCURACY IN ROBUSTNESS TESTS

| Accuracy Indices [p.u.] | | Case 3 | | Case 4 | | Case 5 | |
|---|---|---|---|---|---|---|---|
| | | $Q_1$ | $Q_2$ | $Q_1$ | $Q_2$ | $Q_1$ | $Q_2$ |
| Phase A | Re. Part | 0.020 | 0.018 | 0.020 | 0.017 | 0.019 | 0.017 |
| | Im. Part | 0.014 | 0.013 | 0.014 | 0.012 | 0.013 | 0.014 |
| Phase B | Re. Part | 0.017 | 0.017 | 0.017 | 0.015 | 0.016 | 0.016 |
| | Im. Part | 0.012 | 0.015 | 0.012 | 0.013 | 0.012 | 0.012 |
| Phase C | Re. Part | 0.020 | 0.019 | 0.020 | 0.016 | 0.020 | 0.017 |
| | Im. Part | 0.012 | 0.012 | 0.012 | 0.013 | 0.012 | 0.012 |

TABLE V
COMPARISON IN BALANCED AND UNBALANCED 123-BUS SYSTEMS

| Dimensional Analysis | Balanced | Unbalanced |
|---|---|---|
| State Variables: $[\boldsymbol{x}]$ | 238×1 | 714×1 |
| Measurements and DG Outputs: $[\boldsymbol{z}]$ | 296×1 | 888×1 |
| Augmented Variables: $[\boldsymbol{X}]$ | 534×1 | 1602×1 |
| Coefficient Matrix: $[\boldsymbol{A}]$ | 534×534 | 1602×1602 |

*Case 3*: The intervals of PV and WTG outputs are still [106.72, 149.53] kW and [84.52, 103.31] kW. Their true values are fixed at 107 kW and 85 kW in power flow calculation, respectively, to generate the measurements for ISE.

*Case 4*: True values of PV and WTG outputs are 149 kW and 85 kW, respectively. Other settings are the same as the ones in Case 3.

*Case 5*: True values of PV and WTG outputs are 149 kW and 103 kW, respectively. Other settings are the same as the ones in Case 3.

Table IV summarizes the accuracy indices $Q_1$ and $Q_2$ on three phases in these cases. It is concluded that these estimation results are not significantly affected by the extent of the deviation of DG interval predictions relative to their true values. In contrast, the true values of DG outputs in [13] and [18] are always assumed in the center of their intervals, which may not be robust due to the variability of DG outputs.

## D. *Computational Efficiency*

Numerical experiments are carried out to investigate the computational efficiency of the proposed algorithm. All the tests are performed in MATLAB with the INTLAB toolbox using a 2.5 GHz, 8 GB of RAM, Intel Core i5 computer.

The dimensional analysis towards the ISE model in the balanced and unbalanced 123-bus systems is given in Table V.

TABLE VI
CPU TIME IN DIFFERENT CASES

| CPU Time [s] | Case 1 | | Case 2 | |
|---|---|---|---|---|
| | 13-bus | 123-bus | 13-bus | 123-bus |
| Proposed Method | 0.0099 | 0.847 | 0.012 | 0.965 |
| MC (3000 trials) | 89.52 | 5655.5 | 89.78 | 6209.7 |
| LP | 2.560 | 164.68 | - | - |
| IKO[18] | 8.994 | 795.57 | 9.957 | 911.50 |

TABLE VII
COMPUTATION TIME IN 123-BUS SYSTEM

| Case 2 | Time for Initial Solution [s] | Iteration Number | Average Time in Single Iteration [s] |
|---|---|---|---|
| Proposed Method | 0.876 | 2 | 0.0445 |
| IKO [18] | 908.30 | 8 | 0.400 |

The comparison shows that the unbalanced system leads to a higher-dimensional interval equation. Also, the complexity in unbalanced systems intensifies low efficiency of the existing methods such as [18], which is proposed in balanced systems. Further, in both unbalanced systems, the CPU times of the proposed algorithm, MC simulations, the LP-based method, and the IKO method are summarized in Table VI. It shows that the proposed algorithm solves the ISE model in less than one-hundredth amount of time, compared with other methods.

Algorithmically, the LP-based method requires solving the equal-scale minimum and maximum problems for *n* times, where *n* is the total number of state variables. It should be noted that the LP-based method cannot deal with the uncertain line parameters as in Case 2. Also, as discussed in Section III-C, the initial intervals provided by IGE are much wider than the final solution hull, and the iterative process of the IKO method is time-consuming. The comparative analysis between the proposed algorithm and the IKO method is shown in Table VII. It is concluded that the high computational efficiency of the proposed algorithm firmly holds in the 123-bus unbalanced system.

## E. *Results of Model I and Model III*

We test the proposed algorithm via *Model I* and *Model III* on the 123-bus system, *i.e.*, DG outputs can be obtained by meters or pseudo-measurements. Two cases are designed below.

*Case 6*: Constant line parameter vector, $\boldsymbol{P}_0$. Also, the DG outputs are modeled as pseudo-measurements following known

TABLE VIII
ESTIMATION ACCURACY IN ROBUSTNESS TESTS

| Accuracy Indices [p.u.] | | Case 6 $Q_1$ | Case 6 $Q_2$ | Case 7 $Q_1$ | Case 7 $Q_2$ |
|---|---|---|---|---|---|
| Phase A | Re. Part | 0.0135 | 0.0107 | 0.0136 | 0.0124 |
| | Im. Part | 0.0078 | 0.0090 | 0.0080 | 0.0122 |
| Phase B | Re. Part | 0.0116 | 0.0106 | 0.0120 | 0.0121 |
| | Im. Part | 0.0068 | 0.0093 | 0.0070 | 0.0117 |
| Phase C | Re. Part | 0.0138 | 0.0116 | 0.0139 | 0.0133 |
| | Im. Part | 0.0080 | 0.0084 | 0.0081 | 0.0118 |

TABLE IX
COMPUTATION TIME IN 123-BUS SYSTEM

| CPU Time [s] | Case 6 | Case 7 |
|---|---|---|
| Proposed Method | 0.633 | 0.720 |
| MC (3000 trials) | 5178.2 | 5750.3 |
| LP | 135.57 | - |
| IKO [18] | 300.44 | 316.36 |

Gaussian distributions, and their maximum errors are 10% of active and reactive powers [13].

*Case 7*: Line parameters change in $[0.95\boldsymbol{P}_0, 1.05\boldsymbol{P}_0]$. Other settings are the same as the ones in Case 6.

The indices $Q_1$ and $Q_2$ in both cases are shown in Table VIII. Besides, we list the computation time of these cases in Table IX. The comparison between these two cases illustrates that less uncertainty leads to tighter interval results and higher computational efficiency.

## V. CONCLUSION

A novel and fast ISE algorithm with multiple uncertainties in unbalanced distribution systems is presented. We establish a general ISE framework that simultaneously considers imprecise line parameters, measurements with noises, and uncertain DG outputs. An MKO algorithm is proposed to solve these ISE models and obtain the upper and lower bounds of state variables for better monitoring systems under the coordinated impacts of these multiple uncertainties. The proposed algorithm is tested on unbalanced 13-bus and 123-bus distribution systems. In contrast to MC simulations and the existing alternatives, the proposed algorithm encloses tighter boundaries of state variables in a faster manner. Future work focuses on the applications of the proposed algorithm to bad data detection in active distribution systems with multiple uncertainties.

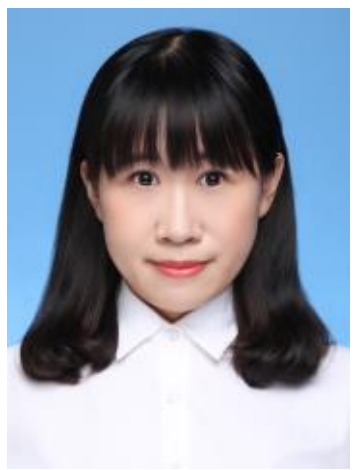

**Ying Zhang** (S’18) received the B.S. and M.S. degrees in electrical engineering from Shandong University, Jinan, China, in 2014 and 2017. She is currently pursuing the Ph.D. degree in Department of Electrical and Computer Engineering at Southern Methodist University, Dallas, Texas, USA. Her research interests include distribution system state estimation and its applications with PMU data in machine learning, fault location, and cyber-security. She won the China National Academic Scholarship for excellent undergraduates twice. She serves as a peer reviewer in IEEE PES letters and Journal of Modern Power Systems and Clean Energy.

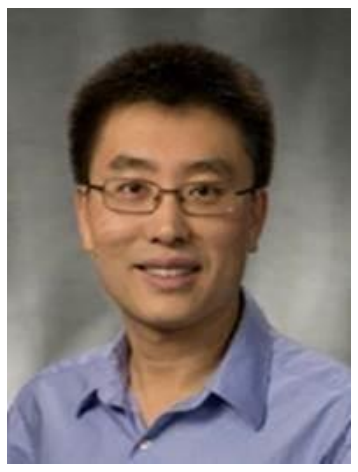

**Jianhui Wang** (M’07-SM’12) received the Ph.D. degree in electrical engineering from Illinois Institute of Technology, Chicago, Illinois, USA, in 2007. Presently, he is an Associate Professor with the Department of Electrical and Computer Engineering at Southern Methodist University, Dallas, Texas, USA. Prior to joining SMU, Dr. Wang had an eleven-year stint at Argonne National Laboratory with the last appointment as Section Lead – Advanced Grid Modeling. Dr. Wang is the secretary of the IEEE Power & Energy Society (PES) Power System Operations, Planning & Economics Committee. He has held visiting positions in Europe, Australia and Hong Kong including a VELUX Visiting Professorship at the Technical University of Denmark (DTU). Dr. Wang is the Editor-in-Chief of the IEEE Transactions on Smart Grid and an IEEE PES Distinguished Lecturer. He is also a Clarivate Analytics highly cited researcher for 2018.

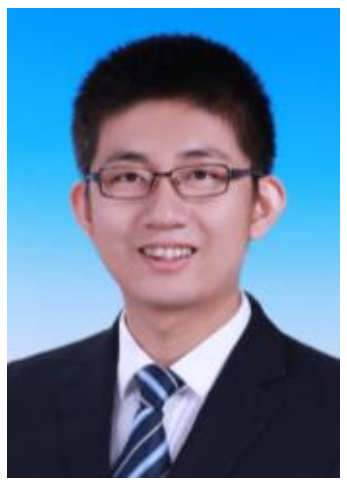

**Zhengshuo Li** (M’16) received his Bachelor and Ph.D. degrees from the Department of Electrical Engineering, Tsinghua University, Beijing, China, in 2011 and 2016. He worked as a postdoctoral fellow at Tsinghua-Berkeley Shenzhen Institute (TBSI) from 2016 to 2018. Then he joined Southern Methodist University as a research assistant professor. His research interests include economic dispatch and security analysis of transmission and distribution grids and demand response in smart grids. He won Special Class Research Fund from China Postdoctoral Science Foundation in 2017 in which year only eight young scholars in electrical engineering were selected for this fund. He also won Best Paper Award of IEEE PES General Meeting and China National Doctoral Academic Annual Meeting, and the Excellent Doctoral Dissertation Award from Tsinghua University in 2016. He was a recipient of the Best Reviewer Award for the IEEE Transactions on Smart Grid and Proceedings of CSEE.